\documentclass[conference]{IEEEtran}
\IEEEoverridecommandlockouts
\usepackage{cite}
\usepackage{amsmath,amssymb,amsfonts}

\usepackage{booktabs}
\usepackage{array}      % for \newcolumntype and \arraybackslash
\usepackage{tabularx}
\newcolumntype{Y}{>{\raggedright\arraybackslash}X} % ragged-right X

\usepackage{booktabs,tabularx,array,ragged2e}
\newcolumntype{Y}{>{\RaggedRight\arraybackslash}X}

\usepackage{algorithmic}
\usepackage{graphicx}
\usepackage{textcomp}
\usepackage{xcolor}
\usepackage{caption}
\usepackage{subcaption}
\usepackage{multirow}
\usepackage{booktabs}
\usepackage{orcidlink}
\def\BibTeX{{\rm B\kern-.05em{\sc i\kern-.025em b}\kern-.08em
    T\kern-.1667em\lower.7ex\hbox{E}\kern-.125emX}}

\usepackage{lipsum}
\usepackage{fancyhdr}
\fancypagestyle{sec}{\lhead{\tmpx}}

\fancypagestyle{arXiv}
{
   \fancyhf{}
   \chead{\textcolor{gray}{This paper has been accepted as a lecture at the 2025 IEEE Biomedical Circuits and Systems Conference (BioCAS), \\Abu Dhabi, UAE, Oct. 16–18, 2025}}
   \cfoot{ \fontsize{=9pt}{10pt}\selectfont  \textcolor{gray}{© 2025 IEEE. Personal use of this material is permitted. Permission from IEEE must be obtained for all other uses, in any current or future media, including reprinting/republishing this material for advertising or promotional purposes, creating new collective works, for resale or redistribution to servers or lists, or reuse of any copyrighted component of this work in other works}\\\fontsize{=10pt}{12pt}\selectfont\thepage}
   
}
    
\begin{document}

% \title{Optimizing OpenAI Whisper for Aphasic Speech Recognition: Fine‑Tuning Strategies and Data‑Ratio Analysis\\}
\title{AS-ASR: A Lightweight Framework for \\Aphasia-Specific Automatic Speech Recognition}
% {\footnotesize \textsuperscript{*}Note: Sub-titles are not captured in Xplore and
% should not be used}
% \thanks{Identify applicable funding agency here. If none, delete this.}

\author{Chen Bao, Chuanbing Huo,
Qinyu Chen\orcidlink{0009-0005-9480-6164
},~\IEEEmembership{Member,~IEEE}, 
Chang Gao\orcidlink{0000-0002-3284-4078},~\IEEEmembership{Member,~IEEE}
\thanks{Chen Bao and Chang Gao are with the Department of Microelectronics, Delft University of Technology, The Netherlands.}
\thanks{Qinyu Chen is with the Leiden Institute of Advanced Computer Science (LIACS), Leiden University, The Netherlands.}
\thanks{Chuanbing Huo is with Sanford Health, Minnesota, United States.}
\thanks{
Corresponding authors: Chang Gao (chang.gao@tudelft.nl) and Qinyu Chen (q.chen@liacs.leidenuniv.nl).
}% <-this % stops a space
}

\maketitle

\begin{abstract}

This paper proposes \textbf{AS-ASR}, a lightweight aphasia-specific speech recognition framework based on Whisper-tiny, tailored for low-resource deployment on edge devices. Our approach introduces a hybrid training strategy that systematically combines standard and aphasic speech at varying ratios, enabling robust generalization, and a GPT–4–based reference enhancement method that refines noisy aphasic transcripts, improving supervision quality. We conduct extensive experiments across multiple data mixing configurations and evaluation settings. Results show that our fine-tuned model significantly outperforms the zero-shot baseline, reducing WER on aphasic speech by over 30\% while preserving performance on standard speech. The proposed framework offers a scalable, efficient solution for real-world disordered speech recognition.
% These improvements demonstrate that fine-tuned AS-ASR provides higher accuracy and low latency and serves as a viable solution for real-time on-device pathological speech recognition.
\end{abstract}


\section{Introduction}
\thispagestyle{arXiv}
Aphasia is a language disorder caused by brain damage due to stroke, traumatic injury, or neurodegenerative disease. Individuals with aphasia experience impairments in both language comprehension and production~\cite{b3}. Their speech is often marked by distorted articulation, hesitations, repetitions, and disrupted fluency. These features deviate significantly from typical speech patterns~\cite{b4}. Such irregularities pose serious challenges for speech-based technologies and highlight the need for automatic speech recognition (ASR) systems that can assist clinicians in assessing, monitoring, and supporting individuals with disordered speech.

Despite recent progress in ASR, mainstream models remain predominantly trained on fluent, well-structured data and struggle to generalize to disordered speech. One such model is Whisper, a versatile encoder decoder ASR framework introduced by OpenAI, trained on 680,000 hours of multilingual and multitask audio data collected from the web~\cite{b1}. While Whisper demonstrates robust performance across a wide range of domains including noisy and accented speech, it, like other standard ASR systems, shows limitations when applied to disordered speech such as aphasia~\cite{b2}. These systems often produce incomplete, incorrect, or misleading transcriptions due to their lack of exposure to disfluent or atypical linguistic patterns during training~\cite{b5}.

In recent years, edge devices have been increasingly integrated into rehabilitation technologies and wearable systems~\cite{b8}. This trend underscores a growing demand for real-time, offline speech recognition capabilities that can operate independently of cloud infrastructure. Edge devices offer several advantages, including low power consumption, portability, cost efficiency, and improved data privacy~\cite{b9}. However, these benefits come at the cost of limited computational resources and memory capacity. As a result, deploying ASR models on such platforms requires careful design choices. Model architecture, size, and inference speed must be optimized to ensure practical and responsive performance~\cite{b10}.

To address these challenges, we propose \textbf{AS-ASR}, a lightweight and aphasia-specific ASR framework based on Whisper, optimized for edge deployment and disordered speech recognition. As illustrated in Fig.~\ref{fig:intro}, AS-ASR can correctly interpret a fragmented aphasic utterance (e.g., “Cuh... I... brown..water”) as “I want a coffee,” whereas a conventional ASR system such as Wav2Vec generates an incomplete or misleading output. This example highlights the importance of domain adaptation and intent recovery when dealing with aphasia speech. The contributions include:
\begin{figure}[tbp]
\centerline{\includegraphics[width=\columnwidth]{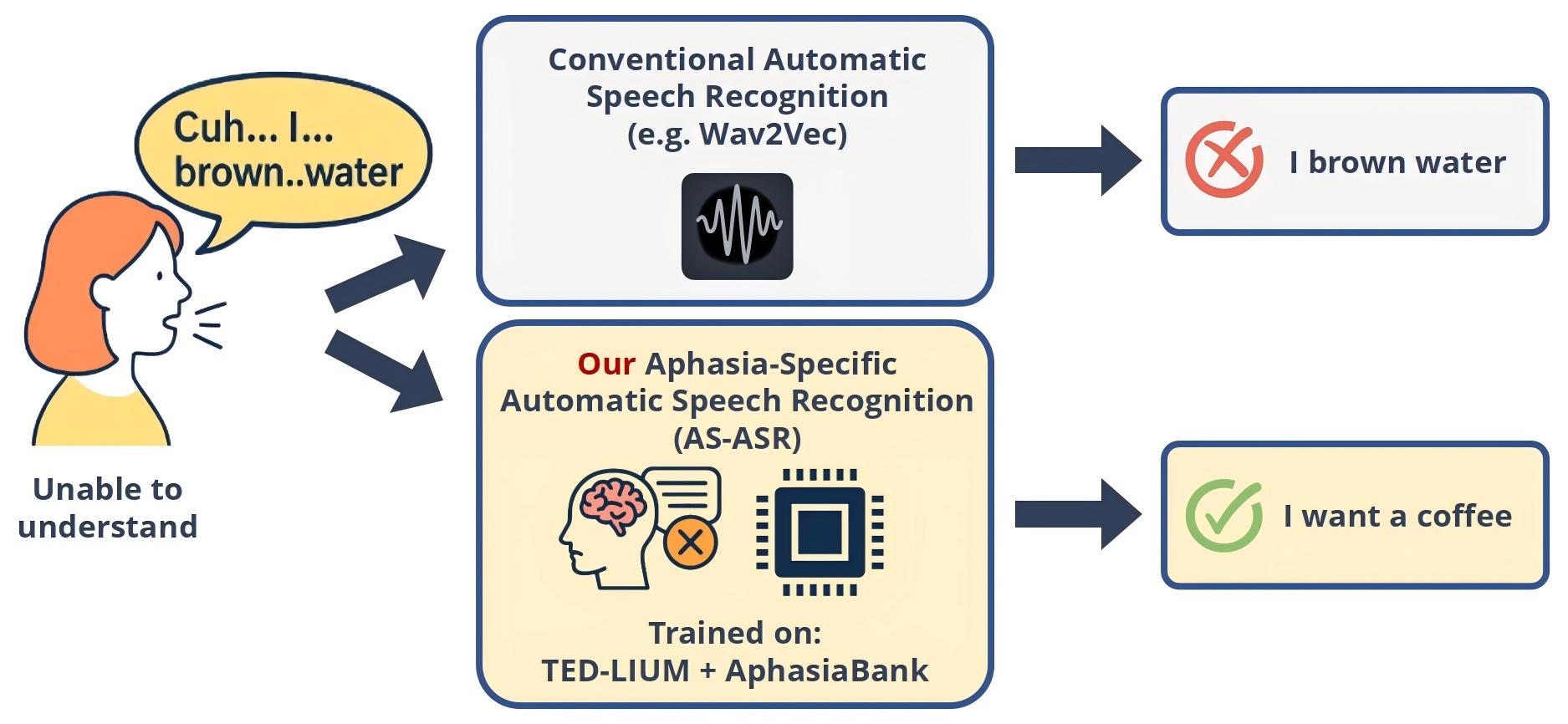}}
\caption{Comparison between conventional and aphasia-specific ASR systems on aphasia speech.}
\label{fig:intro}
\end{figure}

\begin{enumerate}
    \item We propose AS-ASR, a lightweight and aphasia-specific ASR framework based on Whisper, tailored for deployment on edge devices.
    
    \item We construct a hybrid dataset combining pathological and normal speech with varying mixing ratios, enabling systematic analysis of data composition effects on recognition performance.
    
    \item We introduce a GPT-4–based transcript enhancement method that refines noisy aphasic references, improving training quality and model generalization.
\end{enumerate}

\section{Related Work}

Early ASR work on disordered speech sought to distinguish aphasia types. Fraser et al.~\cite{b11} used ASR-derived lexical features for Primary Progressive Aphasia diagnosis despite high word error rates, while Le and Mower Provost~\cite{b12} established an ASR baseline on AphasiaBank, showing small datasets could improve DNN performance via domain adaptation.

More recent work addresses inter-patient speech variability. Perez et al.~\cite{b13} utilized a mixture of experts (MoE) model guided by an intelligibility detector for gains across severity levels. To overcome non-English data scarcity, Chatzoudis et al.~\cite{b14} proposed a zero-shot cross-lingual framework using multilingual models to detect aphasia in Greek and French.

Sanguedolce et al.~\cite{b15} first evaluated Whisper, finding its WER correlates with aphasia severity and showing the limits of models trained only on fluent speech. They later released the SONIVA corpus and achieved state-of-the-art results by fine-tuning Whisper, also analyzing model size effects on generalization~\cite{b23}. These resources, along with Tang et al.’s benchmark~\cite{b16}, provide a reproducible foundation for aphasia ASR.

Separately, Liu et al.~\cite{b17} demonstrated that parameter-efficient fine-tuning achieves comparable gains in low-resource scenarios, enabling the practical deployment of Whisper on clinical and edge devices for users with disordered speech.

\section{Methodology}
This section introduces our aphasia-specific fine-tuning framework, with a focus on baseline model selection, dataset construction, speech reference enhancement, and training data composition strategies.
\subsection{Efficient Model Selection}
This project employs Whisper-tiny-en, the smallest model in the Whisper series with approximately 39 million parameters. As shown in Table~\ref{tab:whisper-models}, this English-only variant is optimized for English tasks, requires only 1GB of VRAM, and offers the fastest inference speed in the series which are about 10 times faster than the large model. These properties make it well-suited for real-time deployment on resource-constrained edge devices such as the Jetson Nano. Despite its compact size, the model achieves notable gains through targeted fine tuning on the acoustic and linguistic features of aphasic speech, making it a strong candidate for low-latency pathological speech recognition.

\begin{table}[htbp]
\centering
\caption{Comparison of Whisper model sizes in terms of parameter count, memory requirements, and relative inference speed.}
\label{tab:whisper-models}
\renewcommand{\arraystretch}{1.2}
\begin{tabular}{lccc}
\toprule
\textbf{Model Size} & \textbf{Parameters} & \textbf{Required VRAM} & \textbf{Relative Speed} \\
\midrule
Tiny   & 39M   & $\sim$1 GB  & $\sim$10$\times$ \\
Base   & 74M   & $\sim$1 GB  & $\sim$7$\times$ \\
Small  & 244M  & $\sim$2 GB  & $\sim$4$\times$ \\
Medium & 769M  & $\sim$5 GB  & $\sim$2$\times$ \\
Large  & 1550M & $\sim$10 GB & 1$\times$ \\
\bottomrule
\end{tabular}
\end{table}

\subsection{Hybrid Data Fine-tuning Strategy}
To enhance the model's adaptability to aphasia speech, we fine-tuned the pre-trained Whisper-tiny-en model using a hybrid training set composed of both aphasic and normal speech data. 
This mixed dataset combines speech samples from individuals with aphasia (AphasiaBank) and normal speakers (TED-LIUM v2), enabling the model to learn both pathological acoustic patterns and standard phonetic structures. By exposing the model to diverse speech characteristics during training, we aim to improve its robustness and generalization in real-world scenarios where speech can vary significantly. The ratio of aphasic-to-normal speech in the training data was systematically varied in subsequent experiments to study its effect on recognition performance.

\subsection{Enhanced Aphasia-Centric Dataset Construction}
For aphasia speech corpus, we utilize speech samples from five sub-corpora of the AphasiaBank database, namely the ACWT Corpus~~\cite{b18}, Adler Corpus~~\cite{b19}, APROCSA Corpus~~\cite{b20}, Boston University Corpus~~\cite{b21}, and University of Kansas Corpus~~\cite{b22}. These corpora provide structured interviews and narrative speech data from individuals with post-stroke aphasia, collected across various clinical centers in the United States. The corpus size is strategically matched to TEDLIUM v2 (around 14,000 segments) to enable controlled cross-corpus comparisons.

Importantly, we apply GPT-4–based reference enhancement to improve the clarity of pathological speech labels while preserving speaker intent in aphasia speech, thereby offering higher-quality supervision for model learning. From the raw .cha transcripts provided by AphasiaBank, we extract patient–clinician dialogue segments with timestamps, remove CHAT-specific markup and non-linguistic annotations, and retain only linguistic content. These utterances are then refined by GPT-4 using prompts that enforce fluent, standard English rewrites without hallucination or paraphrasing. As shown in Table~\ref{tab:gpt4-examples}, GPT-4 enhancements are useful when they improve fluency without changing the original meaning, but may become problematic if they add content not intended by the speaker. These enhanced references are then used to fine-tune the model for improved robustness and intelligibility in recognizing aphasic speech, though such enhancements may risk over-correction that removes diagnostically relevant language features.

\begin{table}[htbp]
\centering
\caption{Examples with and without GPT-4 enhancements}
\label{tab:gpt4-examples}
\renewcommand{\arraystretch}{1.15}
\begin{tabularx}{\columnwidth}{lYY}
\toprule
\textbf{Type} & \textbf{Original} & \textbf{With GPT-4} \\
\midrule
Good & Me go… uh store tomorrow &
I will go to the store tomorrow \\
Bad  & No… want… doctor & I do not want to see the doctor today \\
\bottomrule
\end{tabularx}
\end{table}

\section{Experimental Setup}

\subsection{Data Partitioning}
For each mixing ratio between pathological and normal speech (e.g., 10:90, 50:50), we adopt a consistent partitioning strategy: 11,200 segments for the training set (80\%), 1,400 segments for the validation set (10\%) and 1,400 segments for test set (10\%).
Table~\ref{tab:data-ratios} details the specific composition under different mixing ratios. For instance, in the 10\%:90\% ratio configuration: 1,120 Aphasia training samples and 10,080 TED-LIUM training samples.

\begin{table}[htbp]
\centering
\caption{Data distribution under different Aphasia:TED-LIUM mixing ratios. Each configuration totals 14,000 samples, partitioned into training (80\%), development (10\%), and test (10\%) sets.}
\label{tab:data-ratios}
\renewcommand{\arraystretch}{1.2}
\begin{tabular}{c|ccc|ccc}
\toprule
\textbf{Ratio} & \multicolumn{3}{c|}{\textbf{Aphasia}} & \multicolumn{3}{c}{\textbf{TED-LIUM}} \\
 & \textbf{Train} & \textbf{Dev} & \textbf{Test} & \textbf{Train} & \textbf{Dev} & \textbf{Test} \\
\midrule
10\%:90\% & 1120 & 140 & 140 & 10080 & 1260 & 1260 \\
30\%:70\% & 3360 & 420 & 420 & 7840  & 980  & 980  \\
50\%:50\% & 5600 & 700 & 700 & 5600  & 700  & 700  \\
70\%:30\% & 7840 & 980 & 980 & 3360  & 420  & 420  \\
90\%:10\% & 10080 & 1260 & 1260 & 1120  & 140  & 140  \\
\bottomrule
\end{tabular}
\end{table}

\subsection{Fine-tuning Details}

Fine-tuning was performed on Quechua, a high-performance GPU server equipped with eight NVIDIA RTX A6000 GPUs (48 GB GDDR6 each). The Whisper-tiny model was trained for 10 epochs using a per-device batch size of 16. Mixed-precision (fp16) training and gradient checkpointing were enabled to reduce memory footprint and accelerate throughput. 
The training pipeline was implemented with the HuggingFace \texttt{Trainer} framework, using the AdamW optimizer (learning rate = $5 \times 10^{-5}$) and a linear learning rate scheduler. Model evaluation was conducted after each epoch, with checkpoints saved accordingly.

\subsection{Evaluation Metric}
We adopt the standard Word Error Rate (WER) as the primary evaluation metric:
\begin{equation}
\text{WER} = \frac{S + D + I}{N} \times 100\%
\end{equation}
where S denotes substitutions, D deletions, I insertions, and N the total words in reference transcripts. All evaluations are conducted on the same test sets to ensure comparability.

\subsection{Baseline Systems}

We define two baseline systems for comparative evaluation:

\textbf{Baseline-1} evaluates the zero-shot performance of the pre-trained Whisper-\texttt{Tiny.en} model on three datasets: TED-LIUM only (representing typical speech), Aphasia only (pathological speech), and a merged set combining both. This serves as a reference for the model’s generalization without task-specific adaptation.

\textbf{Baseline-2} reports the average WER of multiple Whisper model variants (Tiny, Base, Small, and Medium) on the pure Aphasia test set. This baseline reflects the inherent challenges of recognizing disordered speech and highlights how model scale affects robustness.

\section{Results \& Analysis}
This section first presents the baseline performance of the untuned Whisper-tiny-en model on the standard TED-LIUM v2 and the aphasic speech corpus to establish reference metrics. Subsequently, we demonstrate the performance of models fine-tuned on TED-LIUM-only and mixed corpora (via separate and merged approaches) to assess the effectiveness of different fine-tuning methodologies. Finally, we systematically analyze the impact of varying aphasia-to-normal speech data ratios on fine-tuning outcomes by comparing model performance across different mixing proportions.

\subsection{Baseline Performance}
Table~\ref{tab:baseline-aphasia} summarizes the average WER of different Whisper model sizes on the pure Aphasia test set (Baseline-2). Among the five Whisper model variants evaluated on the aphasic speech dataset, the Whisper-Large model achieved the lowest average WER of 0.622, demonstrating the best overall recognition accuracy for disordered speech. The \texttt{Medium.en} and \texttt{Small.en} models followed closely, with average WERs of 0.677 and 0.686, respectively, suggesting diminishing returns as model size increases. Surprisingly, the \texttt{Base.en} model performed the worst with a WER of 0.886, even higher than the much smaller \texttt{Tiny.en} model (WER = 0.787). This indicates that larger models do not consistently guarantee better performance in the context of aphasic speech. Given its lightweight architecture (39M parameters), small memory footprint (1GB VRAM), and moderate recognition ability, the Whisper-\texttt{Tiny.en} model remains a promising choice for deployment on low-power, resource-constrained edge devices where real-time inference is required.

\begin{table}[htbp]
\centering
\caption{Baseline-2 WER Results on Aphasia Datasets by different Whisper models.}
\label{tab:baseline-aphasia}
\renewcommand{\arraystretch}{1.2}
\begin{tabular}{lc}
\toprule
\textbf{Model Size} & \textbf{Average WER} \\
\midrule
\texttt{Tiny.en }    & 0.787 \\
\texttt{Base.en  }   & 0.886 \\
\texttt{Small.en }   & 0.686 \\
\texttt{Medium.en }  & 0.677 \\
\texttt{Large}       & 0.622 \\
\bottomrule
\end{tabular}
\end{table}

\subsection{Fine-tuning Results}
Following the baseline evaluation across multiple Whisper variants (Table~\ref{tab:baseline-aphasia}), we selected Whisper-\texttt{Tiny.en} as the foundation for further fine-tuning due to its balance between model size and performance. As shown in Table~\ref{tab:wer_comparison}, we fine-tuned Whisper-\texttt{Tiny.en} on a combined dataset of TED-LIUM and AphasiaBank, and evaluated it under three testing conditions: TED-LIUM only, Aphasia only, and a merged set combining both domains.

Compared to the original Whisper-\texttt{Tiny.en} baseline (Baseline-1 in Table~\ref{tab:wer_comparison}), which operates without any task-specific fine-tuning, the fine-tuned model demonstrates substantial improvements in aphasia and mixed speech. Baseline-1 reflects the model’s zero-shot generalization capability, while it performs well on clean TED-LIUM speech (WER = 0.119/0.117 on dev/test), its performance on aphasic speech deteriorates drastically (WER = 0.808/0.755), and it struggles to generalize under mixed-domain conditions (WER = 0.515/0.498 on the merged set). After fine-tuning on hybrid data, the model retains its recognition capability on fluent speech (WER = 0.116 on both TED-LIUM dev and test), indicating no significant degradation. More importantly, it achieves dramatic gains on aphasic speech, reducing WER to 0.430 (dev) and 0.454 (test). It also shows enhanced robustness on the merged set (WER = 0.162/0.170), underscoring improved domain adaptability.

These results highlight that fine-tuning not only mitigates the baseline model’s limitations on pathological speech but also strengthens its generalization across domains, without sacrificing performance on clean speech. This makes the fine-tuned Whisper-\texttt{Tiny.en} model a strong candidate for deployment in real-time clinical ASR systems operating under diverse acoustic and linguistic conditions.

\begin{table}[htbp]
\centering
\caption{WER of Baseline-1 and Fine-tuned Variants on TED-LIUM and Aphasic Speech}
\label{tab:wer_comparison}
\renewcommand{\arraystretch}{1.2}
\begin{tabular}{llcc}
\toprule
\textbf{Model} & \textbf{Dataset} & \textbf{Dev WER} & \textbf{Test WER} \\
\midrule
\multirow{3}{*}{\shortstack[l]{Whisper-\texttt{Tiny.en} \\ (Baseline-1)}} 
    & TED-LIUM only & 0.119 & 0.117 \\
    & Aphasia only  & 0.808 & 0.755 \\
    & Merged        & 0.515 & 0.498 \\
\midrule
\multirow{3}{*}{\shortstack[l]{Fine-tuned \\ Whisper-\texttt{Tiny.en}}}
    & TED-LIUM only & 0.116 & 0.116 \\
    & Aphasia only  & 0.430 & 0.454 \\
    & Merged        & 0.162 & 0.170 \\
\bottomrule
\end{tabular}
\end{table}

\subsection{Impact of Data Ratios}
\begin{figure}[tbp]
\centerline{\includegraphics[width=0.9\columnwidth]{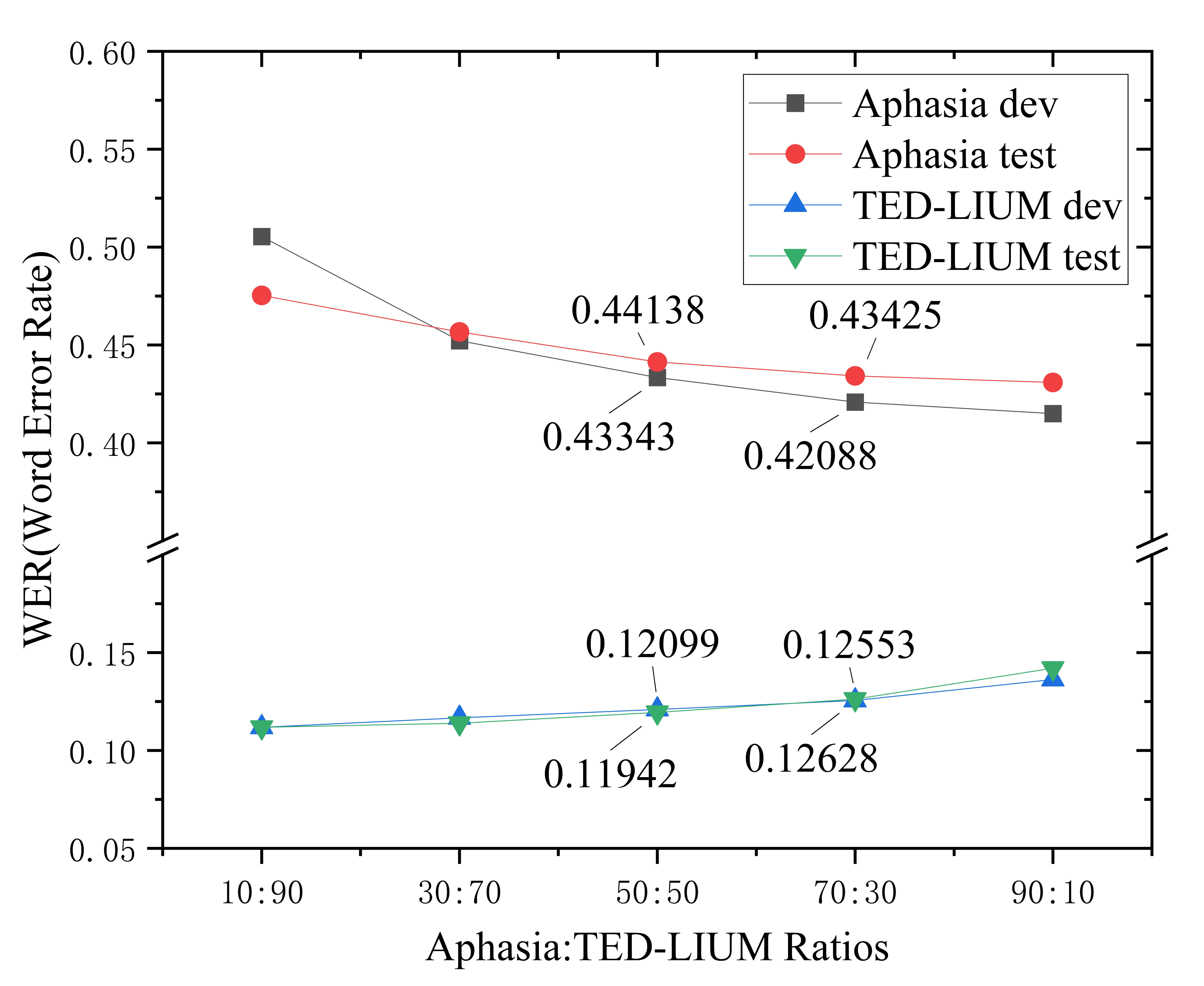}}
\caption{Impact of Aphasia-to-TED-LIUM ratios on WER performance.}
\label{fig:ratios}
\end{figure}

We further investigated how varying the mixing ratios of aphasic and normal speech data (Aphasia:TED-LIUM = 10:90, 30:70, 50:50, 70:30, and 90:10) affects the performance of the fine-tuned Whisper-\texttt{Tiny.en} model. As shown in Fig.~\ref{fig:ratios}, we evaluated the WER on both aphasic and TED-LIUM dev/test sets across different training compositions.

As the proportion of aphasic data increases, a clear trade-off emerges between the two domains:

\begin{itemize}
\item \textbf{Aphasia (black and red lines):}
WER steadily decreases on both dev and test sets as more aphasic speech is introduced during training. This indicates improved adaptation to disordered speech and greater robustness in recognizing pathological utterances.

\item \textbf{TED-LIUM (blue and green lines):}
WER gradually increases, suggesting reduced generalization to clean, fluent speech when the model is increasingly specialized for aphasic input.
\end{itemize}

These findings underscore the importance of data composition in multi-domain ASR. Notably, a balanced configuration (e.g., 50:50 or 70:30) offers a practical compromise, maintaining reasonable recognition performance on both disordered and fluent speech. Such a balance is especially relevant for clinical ASR systems intended to operate across diverse real-world conditions. 

\section{Conclusion \& Future Work}

This work introduced AS-ASR, a lightweight, aphasia-specific speech recognition framework that achieves robust performance at a low computational cost suitable for edge deployment. By leveraging a hybrid dataset of pathological and typical speech, along with GPT-4-enhanced transcripts, our fine-tuned Whisper-based model improves recognition quality for aphasic speech. Future work should include clinician feedback on the usability of the LLM-enhanced transcripts and specialized hardware accelerator design for real product development, drawing inspiration from prior work~\cite{Gao2019,gao2024spartus}, to achieve real-time inference. This would unlock the full potential of AS-ASR in a new generation of clinical and wearable AIoT healthcare products.

\section*{Acknowledgment}
This work was partially supported by the Dutch Research Council (NWO) under the Talent Programme Veni 2023 scheme in Applied and Engineering Sciences (AES), Grant No. 21132 (Energy-Efficient Real-Time Edge Intelligence for Wearable Healthcare Devices), and LIACS Strategic Postdocs and PhD Researchers Program 2024 (Next-Generation LLM System for Post-Stroke Speech Assistance and Rehabilitation). We also thank AphasiaBank for granting access to the aphasic speech dataset, and acknowledge its support by NIH-NIDCD Grant R01-DC008524 (2022–2027).

\newpage
\bibliographystyle{IEEEtran}
\bibliography{IEEEabrv,ref} 

\end{document}